\documentclass[epj]{svjour}

\usepackage{graphicx}
\usepackage{amsmath}
\usepackage{amsbsy}

\newcommand{\TD}{\Theta_{\mathrm{D}}}
\newcommand{\EF}{E_{\mathrm{F}}}
\newcommand{\kB}{k_{\mathrm{B}}}

\begin{document}

\title{Superconducting transition temperatures of the elements related to 
elastic constants}

\author{G. G. N. Angilella\inst{1} 
   \and N. H. March \inst{2,3} 
   \and R. Pucci \inst{1}}

\institute{Dipartimento di Fisica e Astronomia, Universit\`a di 
Catania,\\
and Istituto Nazionale per la Fisica della Materia, UdR di 
Catania,\\
Via S. Sofia, 64, I-95123 Catania, Italy
\and
Department of Physics, University of Antwerp,
Groenenborgerlaan 171, B-2020 Antwerp, Belgium
\and
Oxford University, Oxford, UK}

\date{\today}

\abstract{%
For a given crystal structure, say body-centred-cubic, the many-body 
Hamiltonian $H$ in which nuclear and electron motions are to be treated 
from 
the outset on the same footing, has parameters, for the elements, which 
can be classified as \emph{(i)} atomic mass $M$, \emph{(ii)} atomic number 
$Z$, characterizing the external potential in which electrons move, and 
\emph{(iii)} bcc lattice spacing, or equivalently one can utilize atomic 
volume, $\Omega$.
Since the thermodynamic quantities can be determined from $H$, we conclude 
that $T_c$, the superconducting transition temperature, when it is 
non-zero, may be formally expressed as $T_c = T_c^{(M)} (Z, 
\Omega)$.
One piece of evidence in support is that, in an atomic number \emph{vs} 
atomic volume graph, the superconducting elements lie in a well defined 
region.
Two other relevant points are that \emph{(a)} $T_c$ is related 
by BCS theory, 
though not simply, to the Debye temperature, which in turn is calculable 
from the elastic constants $C_{11}$, $C_{12}$, and $C_{44}$, the atomic 
weight and the atomic volume, and \emph{(b)} $T_c$ for five bcc transition 
metals is linear in the Cauchy deviation $C^\ast = (C_{12} - C_{44} 
)/(C_{12} + C_{44} )$.
Finally, via elastic constants, mass density and atomic volume, 
a correlation between $C^\ast$ and 
the Debye temperature is established for the five bcc transition 
elements.
\PACS{
{74.62.-c}{Transition temperature variations}
\and
{74.70.Ad}{Metals; alloys and binary compounds}
%{74.62.Fj}{Transition temperature variations; pressure effects}
     } % end of PACS codes
} %end of abstract

\authorrunning{G. G. N. Angilella \protect\emph{et al.}}
\titlerunning{Superconducting $T_c$ of elements vs parameters in 
the full Hamiltonian}

\maketitle

\section{Background and outline}

We have recently been concerned with both empirical and theoretical 
relations between the superconducting transition temperature $T_c$ of 
high-$T_c$ cuprates and of heavy Fermion materials 
\cite{Angilella:00b,Angilella:01a,Angilella:03d}.
The generally complex crystallographic structure of such compounds
   makes it difficult to identify useful correlations between their
   superconducting properties (such as $T_c$) and the elastic
   properties of the lattice.
This is not the case of several superconducting elements with a
   definite and relatively simple crystallographic structure,
   \emph{e.g.} characterized by only a few non-zero components of the
   elastic tensor.
Although any such correlation applying to the `simple' superconducting
   elements may not be immediately generalized to other unconventional
   superconductors, they are anyway expected to focus on the relevant
   variables which would be worthwhile studying, both experimentally
   and theoretically, also in the new classes of superconductors.

Following the Bardeen-Cooper-Schrieffer (BCS) theory \cite{Schrieffer:64}
of the metallic elements, firmly rooted in electron-phonon interaction as
the basic mechanism resulting in the formation of Cooper pairs, questions
have come up regarding the role of strong electron-electron interactions
in both the high-$T_c$ cuprates and heavy Fermion systems.

Here, our basic philosophy will be to insist that if we were able to solve
the many-body Schr\"odinger equation for the (considered infinite)  
superconducting materials, then by treating the motion of nuclei and
electrons on the same footing, plus full inclusion of electron-electron
interactions, such uncertainties involved in separating electron-lattice
and Coulomb repulsions between electrons would be bypassed.

Having said that, let us take as the simplest starting point the metallic 
elements. Then, the input information into any computer programme to treat 
these elements would be as follows.
First, of course, we should need to specify the structure.
To be definite, below we shall single out the body-centred cubic (bcc) 
lattice, but everything that follows would be equally applicable to the 
more closely packed face-centred cubic (fcc) structure.
Once the structure is specified, one would need to insert the atomic 
volume $\Omega$ (or, of course, essentially equivalently, the lattice 
parameter $a$).
Then, the external potential created by the nuclei must be specified, 
which requires the atomic number $Z$ as further input.
Since one has a many-body Hamiltonian containing both electron 
and nuclear 
kinetic energies, one needs also the nuclear mass $M$.
Of course, we take as obvious the input additionally of the fundamental 
constants $h$ and $e$, plus the electronic mass $m$.

The conclusion from the many-body Hamiltonian is therefore that, for a 
specified structure which we take to be bcc for reasons that will emerge 
below, the superconducting transition temperature $T_c$, given from the 
many-body partition function once the Schr\"odinger equation has been 
solved depends, apart from the given fundamental constants $h$, $e$ and 
$m$, on $M$, $Z$ and $\Omega$, that is
\begin{equation}
T_c = T_c ^{(M)}(Z, \Omega).
\label{eq:Tc}
\end{equation}
Of course, for all other classes of superconductors than the metallic 
elements we have more than one atomic number, possibly the next 
simplest 
case being the alkali-doped fullerides (see some brief comments 
in 
Section~\ref{sec:conclusions}).

With this as background, the outline of the paper is as follows.
Section~\ref{sec:TcZO} picks out specifically five bcc superconducting 
transition elements, W, Mo, Ta, V and Nb.
Two more elements, Cr and Fe, have low temperature bcc structures but 
exhibit cooperative magnetism at low temperatures (antiferro- and 
ferro-magnetism, respectively) and are not superconductors at the lowest 
temperature they have yet been subjected to.
The five elements listed above are considered in the $(\Omega,Z)$ plane 
with respect to their transition temperatures, the reduced isotope effects 
being taken as evidence that in Eq.~(\ref{eq:Tc}) there is, at most, a 
weak and therefore relatively unimportant dependence of $T_c$ on nuclear 
isotopic mass.
Since even then $T_c = T_c^{(M)} (Z,\Omega)$ presents problems 
in its 
representation, 
Section~\ref{sec:Cauchy} reintroduces a classification of the above five 
elements in which $T_c$ is related to the Cauchy discrepancy, \emph{i.e.} 
the departure of $C_{12}$ from $C_{44}$, where these are two of the three 
elastic constants ($C_{11}$ being the other) required to characterize a 
cubic crystal.
Section~\ref{sec:Debye} then returns to an essential ingredient 
of BCS 
theory, and by using a semiempirical approach throws light on the way the 
Cauchy deviation relates to the Debye temperature.
Section~\ref{sec:conclusions} constitutes a summary, plus some proposals 
for further studies, both theoretical and experimental, which should prove 
fruitful.
An Appendix considers zero temperature properties, and in 
particular critical field $H_c (0)$ and energy gap $E_g (0)$, as 
functions of the Cauchy discrepancy.

\section{Dependence of $T_c$ on atomic number $Z$ and atomic volume 
$\Omega$ in bcc transition elements}
\label{sec:TcZO}

\begin{figure}[tb]
\centering
\includegraphics[height=0.9\columnwidth,angle=-90]{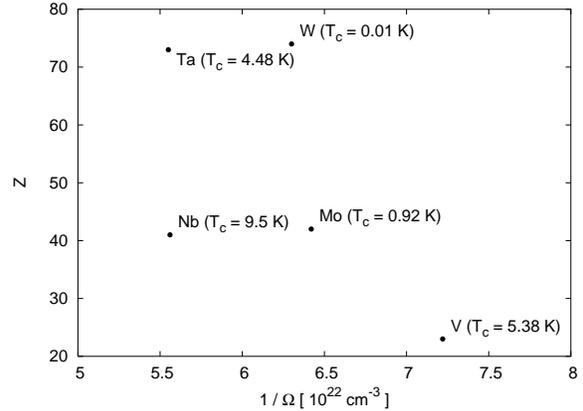}
\caption{%
Shows five superconducting bcc transition elements in the
$(\Omega,Z)$ plane.  Actual abscissa is the reciprocal of the
atomic volume, $\Omega^{-1}$. To each point, the value of the
critical temperature $T_c$ is attached. 
}
\label{fig:OZ}
\end{figure}

In Fig.~\ref{fig:OZ}, a plot is made of the positions of the 
five elements in the ($\Omega,Z)$ plane, the values of $T_c$ 
being attached to these coordinates.

\begin{table}[tb]
\caption{Pressure derivatives of $T_c$ \protect\cite{Poole:95},
   experimental bulk moduli \protect\cite{Mehl:96}, and inferred
   partial derivatives of $T_c$ with respect to $\Omega$ at constant
   $Z$, Eq.~(\protect\ref{eq:dTcdO}), of the bcc superconducting
   transition metals at $P=0$.} 
\centering
\begin{tabular}{|c|ccccc|}
\hline
Element & W & Mo & Ta & V & Nb \\
\hline
$\displaystyle\frac{\partial T_c}{\partial P}$ [K/GPa] 
& --- & $-1.4$ & $-2.6$ & $6.3$ & $-2.0$ \\
$B$ [GPa] & 323 & 272 & 200 & 162 & 170 \\
$\displaystyle\frac{\partial T_c}{\partial\Omega}$ [$10^{31}
   \mathrm{K}\cdot\mathrm{m}^{-3}$] & --- & 2.4 & 2.9 & $-7.4$ & 1.9 \\
\hline
\end{tabular}
\label{tab:dTcdP}
\end{table}

That both $\Omega$ and $Z$ are important variables in 
characterizing $T_c$ is immediately apparent.
As to the functional form $T_c (\Omega,Z)$, one can comment 
that:
\emph{(i)} For constant atomic volume, $T_c$ markedly decreases 
with increasing atomic number.
\emph{(ii)} For constant $Z$, there is plainly substantial 
variation of $T_c$ with atomic volume, which is proportional to 
the reciprocal of the concentration.
Relevant to such variation is the pressure dependence of $T_c$ 
for a given element, provided one remains within the bcc phase.

Despite high pressure can turn many elements into superconductors via
   an insulator-metal transition, $T_c$ usually decreases with
   increasing pressure for most superconducting elements at ambient
   pressure (see Table~\ref{tab:dTcdP}).
Within BCS theory [see also Eq.~(\ref{eq:BCS}) below] or its extension
   by McMillan, this is usually
   justified in terms of a pressure-induced lattice stiffening, which
   reduces the electron-phonon constant at a more rapid rate than the
   electron density of states at the Fermi level is increased
   \cite{Schilling:01}.
Pressure derivatives of $T_c$ can then be straightforwardly related to
   volume derivatives (at constant $Z$) from the relation
\begin{equation}
\frac{\partial \log T_c}{\partial \log \Omega} = - B \frac{\partial
   \log T_c}{\partial P} ,
\label{eq:dTcdO}
\end{equation}
where $B$ is the bulk modulus (see Table~\ref{tab:dTcdP}).

However, even given some knowledge of these partial derivatives, 
the fact that $T_c$ depends apparently in a sensitive way on 
these two variables for the chosen bcc structure leaves open the 
detailed form of the function $T_c (\Omega, Z)$ for this 
structure.
Therefore in the following section we appeal to a known, but so 
far rather neglected, correlation between $T_c$ and the Cauchy 
discrepancy $C_{12} - C_{44}$ between elastic constants.
This is important for our present study, since it is clear that 
$T_c$ can, in fact, be characterized by a single variable, 
rather than the pair $(\Omega,Z)$ used in Fig.~\ref{fig:OZ}.

\section{Characterization of $T_c$ by the Cauchy discrepancy for the five 
bcc transition elements}
\label{sec:Cauchy}

\begin{figure}
\centering
\includegraphics[height=0.9\columnwidth,angle=-90]{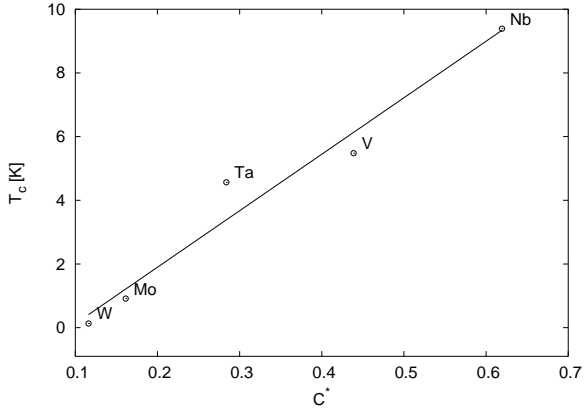}
\caption{Superconducting critical temperatures for bcc 
transition elements correlated with Cauchy discrepancy, 
Eq.~(\protect\ref{eq:Cauchy}).
Solid line is best fit to the points
[see text for discussion, in particular 
Eq.~(\protect\ref{eq:Ledbetter})].
Redrawn from 
Ref.~\protect\cite{Ledbetter:80}
(see also \protect\cite{Mehl:96} for an update on elastic 
constants data).} 
\label{fig:Ledbetter}
\end{figure}

Fig.~\ref{fig:Ledbetter}, redrawn from the work of Ledbetter 
\cite{Ledbetter:80} carried out almost a quarter of a century 
ago, shows a plot of $T_c$ versus the quantity $C^\ast$ defined 
by
\begin{equation}
C^\ast = \frac{C_{12} - C_{44}}{C_{12} + C_{44}} .
\label{eq:Cauchy}
\end{equation}
Ledbetter \cite{Ledbetter:80} also included some alloys, namely 
Nb$_{0.9}$Zr$_{0.1}$, Nb$_{0.4}$Ti$_{0.6}$, and 
Ti$_{0.7}$V$_{0.3}$, but we have omitted these from the redrawn 
Fig.~\ref{fig:Ledbetter}, even though the alloys support the 
general trend of the correlation shown.
Also, the points for the bcc elements Cr and Fe have been
   omitted, since these elements are both characterized by magnetic
   order and no 
   superconductivity in normal conditions.
It should be mentioned, however, that a high-pressure, non-magnetic,
   but also a non-bcc phase of 
   iron has been recently reported to display superconductivity with
   $T_c < 2$~K between 15 and 30~GPa \cite{Shimizu:01}.

The equation of the straight line drawn in Fig.~\ref{fig:Ledbetter} is
\begin{equation}
T_c ~ \mathrm{[K]} = A C^\ast - B,
\label{eq:Ledbetter}
\end{equation}
with $A=17.7$~K and $B=1.65$~K.
Though presently we do not have theory to allow the evaluation 
of $A$ and $B$ from first principles, the correlation in
Eq.~(\protect\ref{eq:Ledbetter}) leads us, in the following 
section, to attempt to relate $C^\ast$ to a basic ingredient of 
BCS theory, the Debye temperature $\TD$.

\section{Cauchy discrepancy related to $\TD$, which gives the 
`scale' of $T_c$ in the BCS theory}
\label{sec:Debye}

\begin{figure}[tb]
\centering
\includegraphics[height=0.9\columnwidth,angle=-90]{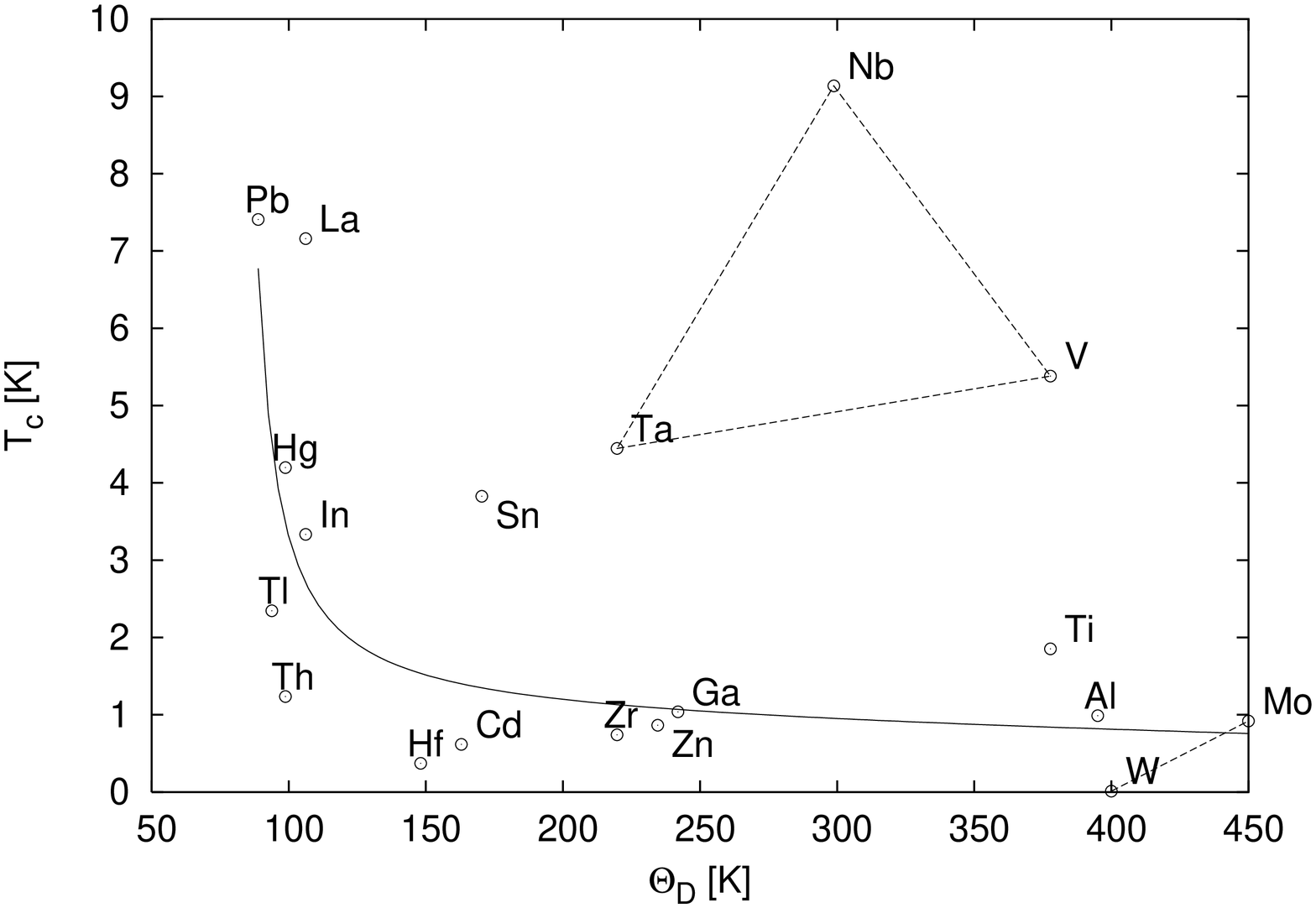}
\includegraphics[height=0.9\columnwidth,angle=-90]{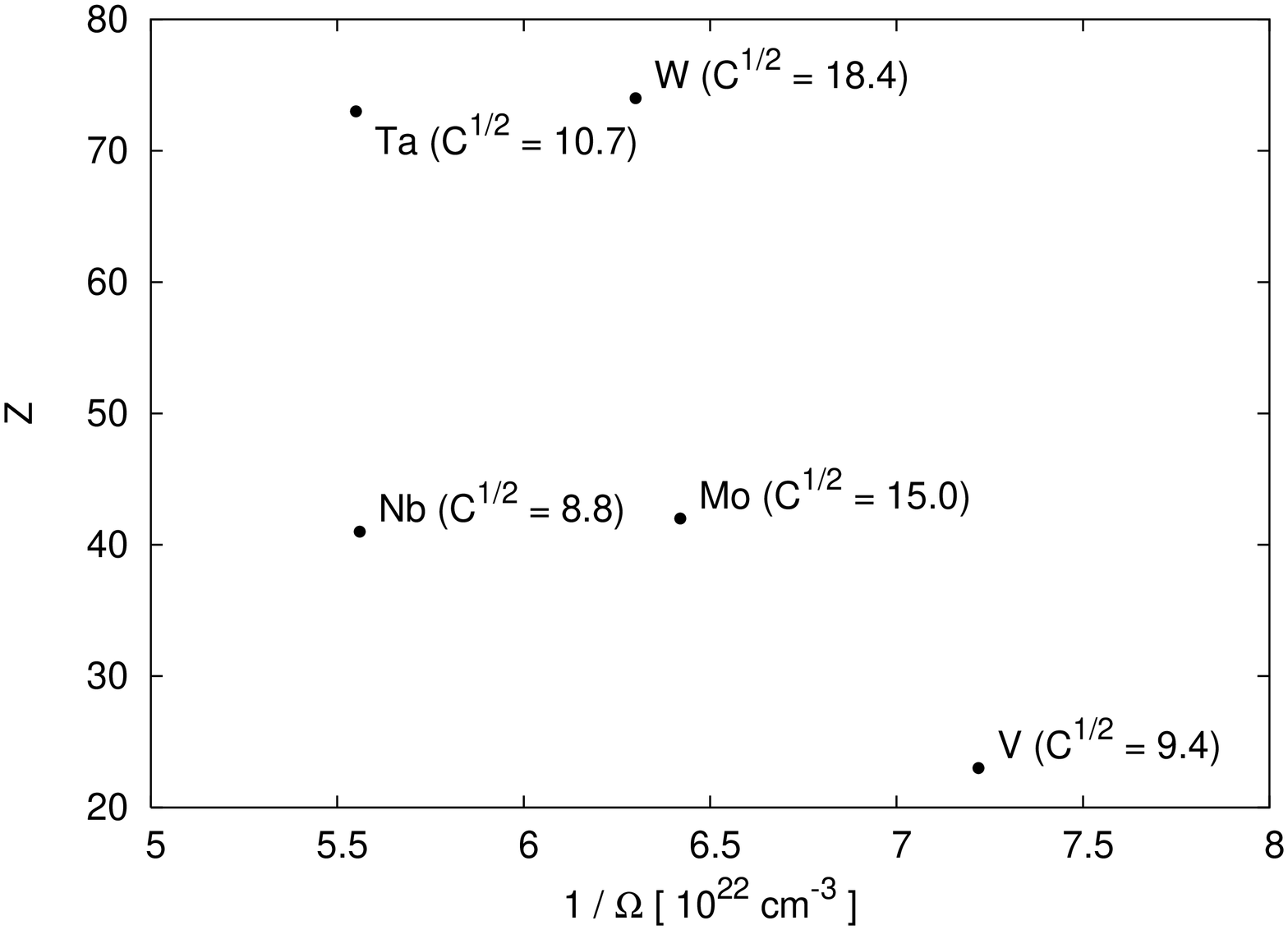}
\caption{%
\underline{\sl Top:}
Superconducting transition temperatures versus Debye 
temperatures for numerous elements.
Redrawn after Ref.~\protect\cite{deLaunay:47}.
Solid and dashed lines are guides to the eye.
See text for discussion.
\underline{\sl Bottom:}
Same as Fig.~\protect\ref{fig:OZ}, but now with attached values 
of 
square root $C^{\frac{1}{2}}$ of the `average' elastic constant,
as defined by Eq.~(\protect\ref{eq:effC}), in GPa$^{\frac{1}{2}}$.
}
\label{fig:deLaunay}
\end{figure}

As Allen and Mitrovic \cite{Allen:82} have stressed, 
notwithstanding the numerous impressive and successful 
predictions for the metallic elements of BCS theory, their 
formula (see Ref.~\cite{Allen:82}, Eq.~(2.29))
\begin{equation}
T_c = 1.13 \TD \exp \left( - \frac{1}{N(\EF)V} \right),
\label{eq:BCS}
\end{equation}
where $N(\EF)$ is the density of states at the Fermi level and 
$V$ is the electron-phonon coupling constant,
is not a successful way of correlating values of $T_c$ for the 
metallic elements.
Nevertheless, it suggests that one should re-open empirically 
the question of a correlation between $T_c$ and $\TD$.
Therefore, more generally than for the bcc transition metals, we 
have redrawn data by de~Launay and Dolecek \cite{deLaunay:47} in 
Fig.~\ref{fig:deLaunay} (top panel), adding values for W and Mo.

While, for mainly non-transition elements, the continuous line 
drawn in Fig.~\ref{fig:deLaunay} (top panel), already given by 
de~Launay and 
Dolecek \cite{deLaunay:47} a decade before BCS theory, shows a 
relation between $T_c$ and $\TD$, it is far from simple.
And the triangle involving Ta, V and Nb 
modified from the 1947 figure of de~Launay and Dolecek 
\cite{deLaunay:47} shows no relation to the continuous curve.

Nevertheless, $\TD$ lies deeply enough in first principles 
theory to enquire whether it can be connected, albeit not 
simply, with the Cauchy deviation $C^\ast$, which is much more 
directly related to $T_c$, as shown in the previous section.

\begin{figure}[tb]
\centering
\includegraphics[height=0.9\columnwidth,angle=-90]{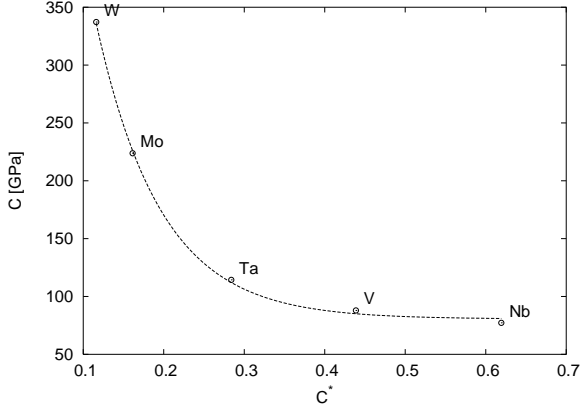}
\caption{Shows `average' elastic constant $C$, 
Eq.~(\protect\ref{eq:effC}), as a function of Cauchy discrepancy 
$C^\ast$, Eq.~(\protect\ref{eq:Cauchy}), for the five bcc 
transition metals.
Dashed line is a guide for the eye, which actually employs an
exponential functional form, of the kind $C = \alpha + \beta 
\exp (-\gamma C^\ast )$, with $\alpha=80.5$~GPa, $\beta=1008.9$~GPa, 
$\gamma = 12.5$. }
\label{fig:average}
\end{figure}

To attempt this, we note that numerous earlier workers have 
calculated the Debye temperature for cubic crystals from 
knowledge of the elastic constants $C_{11}$, $C_{12}$ and 
$C_{44}$, the mass density and the atomic volume $\Omega$.
While Houston's method \cite{Houston:48} is favoured, and has 
been 
developed by Betts \emph{et al.} \cite{Betts:56a,Betts:56b}, we 
have found the semi-empirical relation of Blackman 
\cite{Blackman:51}, quoted in Huntington's review article 
\cite{Huntington:58}, a useful starting point.
This reads
\begin{eqnarray}
\TD^3 &=& \frac{3.15}{8\pi} \left( \frac{h}{\kB} \right)^3
\frac{s}{\rho^{\frac{3}{2}} \Omega} \nonumber \\
&&\times
(C_{11} - C_{12} )^{\frac{1}{2}}
( C_{11} + C_{12} + 2 C_{44} )^{\frac{1}{2}} 
C_{44}^{\frac{1}{2}} ,
\label{eq:Blackman}
\end{eqnarray}
where $s$ is the number of atoms in the unit cell
and $\rho$ is the mass density.
This approximate result, Eq.~(\ref{eq:Blackman}), motivates the 
definition of an `average' elastic constant
\begin{equation}
C = \left( \frac{8\pi}{3.15} \right)^{\frac{2}{3}}
\left( \frac{\kB}{h} \right)^2 \frac{\rho 
\Omega^{\frac{2}{3}}}{s^{\frac{2}{3}}} \TD^2 ,
\label{eq:effC}
\end{equation}
and Fig.~\ref{fig:deLaunay} (bottom panel) parallels 
Fig.~\ref{fig:OZ} 
except that coordinates in the $(\Omega,Z)$ plane are now 
labelled by $C^{\frac{1}{2}}$.
Evidently from this figure at constant volume $\Omega$, 
$C^{\frac{1}{2}}$ related to $\TD$ through Eq.~(\ref{eq:effC}) 
increases with increasing $Z$, in contrast to the behaviour of 
$T_c$ in Fig.~\ref{fig:OZ}.
Also at constant $Z$, $C^{\frac{1}{2}}$ increases with 
decreasing atomic volume.
Nevertheless, again prompted by the BCS theory, we have sought 
to correlate $C^{\frac{1}{2}}$ with $T_c$, but now via the 
Cauchy discrepancy.
Fig.~\ref{fig:average} shows, for the five bcc elements, that 
there is indeed a marked correlation, the functional form 
obtained empirically being recorded in the caption.

\section{Summary and directions for future work}
\label{sec:conclusions}

Though the top panel of Fig.~\ref{fig:deLaunay} makes it quite 
clear that there is no simple relation between Debye temperature 
$\TD$ and the superconducting temperature $T_c$, we have been 
led, via the considerations from Fig.~\ref{fig:OZ} and 
Fig.~\ref{fig:deLaunay} (lower panel), to attempt to correlate 
$C$, having dimensions of an elastic constant and defined in 
Eq.~(\ref{eq:effC}), which in turn involved $\TD$, with the 
Cauchy discrepancy $C^\ast$ in Eq.~(\ref{eq:Cauchy}).
These quantities, for the five bcc elements we have focussed on 
here, are clearly inter-related, as Fig.~\ref{fig:average} 
demonstrates, and the functional form has been extracted.
Since, as Ledbetter \cite{Ledbetter:80} already pointed out in 
1980, $T_c$ relates linearly to $C^\ast$ as in 
Fig.~\ref{fig:Ledbetter}, there is a clear correlation between 
$T_c$ and $\TD$, with mass density and atomic volume entering 
through the definition of the `average' elastic constant in 
Eq.~(\ref{eq:effC}).
Furthermore, and again motivated by BCS theory, zero temperature 
quantities, namely critical field $H_c (0)$ and energy gap $E_g 
(0)$ are shown also to correlate simply with the Cauchy 
discrepancy $C^\ast$ for the five bcc superconducting transition 
elements in Figs.~\ref{fig:HcEg} and \ref{fig:ratio}.

The present work stimulates thoughts concerning generalization 
of the basic approach set out here to other groups of 
superconductors.
Our view is that the next simplest class to study is the 
alkali-doped C$_{60}$ compounds, the fullerides, which have been 
reviewed by Gunnarsson \cite{Gunnarsson:97}.
In Fig.~3 of this review, Gunnarsson has plotted $T_c$ for 
Rb$_3$C$_{60}$ and 
K$_3$C$_{60}$, as a function of lattice parameter, which was 
varied by applying pressure, $P$.
There is a remarkably linear increase in $T_c$ with $P$.
For Na$_2$Rb$_x$Cs$_{1-x}$C$_{60}$ there is again a linear 
variation of $T_c$ with lattice parameter, but with a much 
steeper slope.
It will be of interest for the future to attempt a 
generalization of the approach given here for the metallic 
elements to the fullerides, in view of these striking 
correlations between $T_c$ and lattice parameter.

\begin{acknowledgement}
G.G.N.A. thanks the Department of Physics, University of Antwerp, 
for much hospitality.
\end{acknowledgement}

\appendix

\section{Zero temperature properties, related by BCS theory to 
$T_c$, as functions of Cauchy discrepancy $C^\ast$}

\begin{figure}[tb]
\centering
\includegraphics[height=0.9\columnwidth,angle=-90]{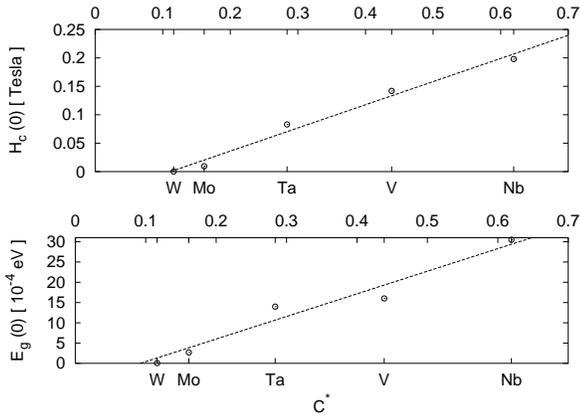}
\caption{Experimental results for critical field $H_c (0)$ and 
energy gap $E_g (0)$, as extracted from tunnelling 
measurements, \emph{vs} Cauchy discrepancy $C^\ast$, as defined 
in Eq.~(\protect\ref{eq:Cauchy}).
} 
\label{fig:HcEg} 
\end{figure}

The purpose of this Appendix is to display a marked correlation 
between experimentally estimated values of the critical field 
$H_c (0)$ and the energy $E_g (0)$, as extracted from tunnelling 
experiments, and the Cauchy discrepancy $C^\ast$, 
Eq.~(\ref{eq:Cauchy}).
Thus, the upper panel of Fig.~\ref{fig:HcEg} shows $H_c (0)$ 
plotted against $C^\ast$ for the five bcc transition elements on 
which attention was focussed in the body of the text.
The data for $H_c (0)$ have been taken from Kittel's book 
\cite{Kittel:96} (see also \cite{Myers:97}).
Evidently, as for $T_c$ in Fig.~\ref{fig:Ledbetter}, a marked 
correlation again exists, the dashed line having the equation
\begin{equation}
H_c (0) = b_1 C^\ast - b_2 ,
\label{eq:Hcfit}
\end{equation}
with $b_1 = 0.41$~Tesla and $b_2 = 0.05$~Tesla.

The energy gap $E_g (0)$, as estimated from tunnelling 
experiments, has again been taken from data given by Kittel 
\cite{Kittel:96} (see also \cite{Poole:95}), and is shown in the 
lower panel of 
Fig.~\ref{fig:HcEg} for the same set of bcc transition elements.
The dashed line, but now with more substantial scatter of the 
experimental points than for either $T_c$ or $H_c (0)$, has the 
equation
\begin{equation}
E_g (0) = e_1 C^\ast - e_2 ,
\label{eq:Egfit}
\end{equation}
with $e_1 = 55.8\cdot 10^{-4}$~eV and $e_2 = 5.16 \cdot 
10^{-4}$~eV.

\begin{figure}
\centering
\includegraphics[height=0.9\columnwidth,angle=-90]{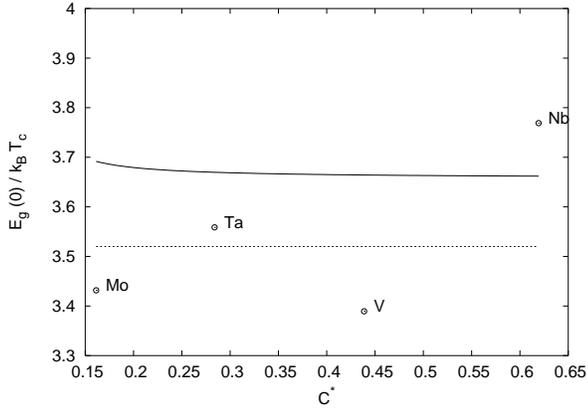}
\caption{BCS ratio $E_g (0) /\kB T_c$ for bcc transition 
elements.
Solid line is ratio of Eqs.~(\protect\ref{eq:Ledbetter}) and 
(\protect\ref{eq:Egfit}), while dashed line is BCS theoretical 
value.}
\label{fig:ratio}
\end{figure}

BCS theory predicts a constant value for the ratio $E_g (0)/\kB 
T_c$. Since Eq.~(\ref{eq:Ledbetter}) and (\ref{eq:Egfit}),
respectively, give a reasonable overall fit for the five bcc 
transition elements focussed on in the present study, we have 
finally plotted the `average' ratio $E_g (0)/\kB
T_c$ from Eqs.~(\ref{eq:Ledbetter}) and (\ref{eq:Egfit}) as a 
function of $C^\ast$ in Fig.~\ref{fig:ratio} using the given 
parameters $A$, $B$, $e_1$ and $e_2$.

\bibliographystyle{mprsty}
\bibliography{Angilella,a,b,c,d,e,f,g,h,i,j,k,l,m,n,o,p,q,r,s,t,u,v,w,x,y,z,zzproceedings}

\end{document}